\documentclass[prl,twocolumn,showpacs,superscriptaddress]{revtex4}

\usepackage{graphicx, natbib}
\begin{document}

\title{High-contrast dispersive readout of a superconducting flux qubit \\ using a nonlinear resonator}

\author{A. Lupa\c scu}
\affiliation{Kavli Institute of Nanoscience, Delft University of
Technology, PO Box 5046, 2600GA Delft, The Netherlands}
\author{E.F.C. Driessen}
\affiliation{Kavli Institute of Nanoscience, Delft University of
Technology, PO Box 5046, 2600GA Delft, The Netherlands}
\author{L. Roschier}
\affiliation{Low Temperature Laboratory, Helsinki University of Technology, PO Box 2200, FIN-02015 HUT, Finland}
\author{C.J.P.M. Harmans}
\affiliation{Kavli Institute of Nanoscience, Delft University of
Technology, PO Box 5046, 2600GA Delft, The Netherlands}
\author{J.E. Mooij}
\affiliation{Kavli Institute of Nanoscience, Delft University of
Technology, PO Box 5046, 2600GA Delft, The Netherlands}

\date{ \today}

\begin{abstract}

We demonstrate high-contrast state detection of a superconducting flux qubit. Detection is realized by probing the microwave transmission of a nonlinear resonator, based on a SQUID. Depending on the driving strength of the resonator, the detector can be operated in the monostable or the bistable mode.  The bistable operation combines high-sensitivity with intrinsic latching. The measured contrast of Rabi oscillations is as high as 87 $\%$; of the missing 13 $\%$, only 3 $\%$ is unaccounted for. Experiments involving two consecutive detection pulses are consistent with preparation of the qubit state by the first measurement. 

\end{abstract}
\maketitle

Superconducting qubits~\cite{_devoret_2004_1} have been established as promising candidates for the implementation of a quantum information processor~\cite{nielsen_2000_1}. Remarkable achievements in this field include the realization of complex single-qubit manipulation schemes~\cite{collin_2004_1} and the generation of entangled two qubit states~\cite{yamamoto_2003_1, mcdermott_2005_1}.

For qubit readout, several detectors have been investigated experimentally. In general, their efficiency is relatively poor, for reasons which are presently not well understood. Most detectors to date rely on irreversible processes in mesoscopic Josephson circuits~\cite{vion_2002_1,chiorescu_2003_1,duty_2004_1,cooper_2004_1,astafiev_2004_1}. For these schemes, energy is dissipated on the chip where the qubit is placed. Therefore long waiting times are necessary to bring the qubit, readout, and control circuits to their proper initial state. In particular the qubit state is strongly disturbed. More recently, \emph{dispersive} measurement schemes are being investigated, which overcome these drawbacks~\cite{grajcar_2004_1,lupascu_2004_1,wallraff_2005_1,_siddiqi_2005_1,sillanpaa_2005_1}. They are based on the measurement of the impedance of a resonator coupled to the qubit. The energy used to probe this resonator is mostly dissipated at a place remote from the qubit chip. Also, the backaction on the qubit is low. However, the qubit relaxation times are typically comparable to the time necessary to for a reliable measurement of the impedance. This limits the detection efficiency when a linear resonator is used~\cite{lupascu_2004_1,wallraff_2005_1}. With a nonlinear resonator, this limitation can be removed, by using a bifurcation transition~\cite{siddiqi_2004_1}. In this case, the result of a qubit measurement is either of two possible oscillation states of the driven resonator. These oscillation states can be latched and reliably discriminated irrespective of subsequent relaxation of the qubit. A dispersive method using latching was successfully demonstrated in~\cite{_siddiqi_2005_1}, where it was used for the readout of a quantronium. 

In this letter we present experimental results on the readout of a superconducting flux qubit using a dispersive method. We observed coherent oscillations in two distinct operation modes of the detector: monostable for weak driving and bistable for strong driving of our non-linear resonator. In the bistable regime the measurement contrast is very large, 87~$\%$, which is a significant improvement over previous measurements~\cite{vion_2002_1,chiorescu_2003_1,duty_2004_1,cooper_2004_1,astafiev_2004_1,_siddiqi_2005_1, wallraff_2005_1}. We also performed consecutive measurements of the qubit. We find that the results are consistent with projection of the qubit state by a first measurement, with a probability as high as 90~$\%$.

\begin{figure}[!]
\includegraphics[width=3.4in]{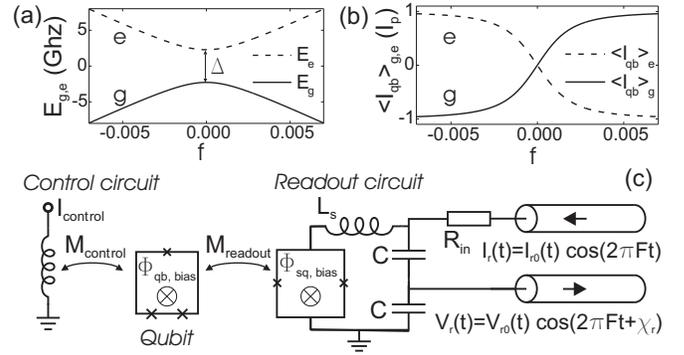}
\caption{\label{fig1} Energy (a) and current (b) of the two qubit states versus applied magnetic flux. (c) Schematic diagram of the qubit and its control and readout circuits. Crosses represent Josephson junctions. The probe wave is sent through a bias resistor $R_{in}=$ 4.7 k$\Omega$. The values of the other circuit components are ${L_{s}\approx 0.2}$ nH, ${C=67}$ pF, ${M_{control}\approx 15}$ fH, and ${M_{readout} =15}$ pH.}
\end{figure}

In our experiments we use a persistent current qubit (PCQ)~\cite{mooij_1999_1}. It is formed of a superconducting ring interrupted by three Josephson junctions. For suitable parameters of the Josephson junctions and for an applied magnetic flux in the qubit loop $\Phi_{qb}$ close to $(n+1/2)\Phi_{0}$ ($n$ is integer and $\Phi_{0}=h/2e$), this system can be operated as a two level system. The qubit Hamiltonian is
\begin{equation}\label{eq_Ham}
  H_{qb}={\textstyle\frac{1}{2}}(\epsilon\sigma_{\!z} + \Delta\sigma_{\!x}) \;,
\end{equation}
where $\epsilon=2I_{p}\Phi_{0}f$. $\Delta$ and $I_{p}$ are fixed parameters, determined by the charging and Josephson energies of the Josephson junctions, and the frustration $f=( \Phi_{qb}-(n+1/2)\Phi_{0}) / \Phi_{0}$. Figure~\ref{fig1}a shows a plot of the  ground (g) and excited (e) energy levels of the qubit used in our measurements. The expectation value of the circulating current in the qubit ring corresponding to each eigenstate is plotted in Fig.~\ref{fig1}b.

Figure~\ref{fig1}c shows a schematic representation of our readout and control circuit. The readout circuit is based on a DC-SQUID, coupled to the qubit through their mutual inductance $M_{readout}$. The SQUID has a Josephson inductance~\cite{barone_1982_1} $L_{J}(\Phi_{sq})$, which depends on the total magnetic flux in its loop $\Phi_{sq}=\Phi_{sq, bias}+M_{readout}I_{qb}$; $\Phi_{sq, bias}$ and $M_{readout}I_{qb}$ are generated by an external coil and by the qubit, respectively. The contribution of the self-generated flux is negligible compared to $\Phi_{sq, bias}$ and nearly constant over the range of variation of $\Phi_{sq,bias}$ relevant in our experiments. The two qubit states have different currents $I_{qb}$ (see Fig.~\ref{fig1}b), resulting in values of the Josephson inductance different by $\approx$ 2 $\%$ in our experiment. We form a resonant circuit by including $L_{J}$ in a loop closed by the capacitors $C$ and the small unintended stray inductance $L_{s}$ (see Fig.~\ref{fig1}c). We probe the transmission of this resonant circuit by sending a wave with a frequency $F$ and current amplitude at the sample $I_{r,0}$ and measuring the amplitude $V_{r,0}$ and relative phase $\chi$ of the voltage of the transmitted wave. We use a low-noise cryogenic amplifier, described in~\cite{roschier_2004_1}. When $F$ is close to the resonance frequency $F_{res}$, the transmission depends strongly on small changes of $L_{J}$ and thus on the qubit state. In our experiment $F_{res}\approx$ 800 MHz and the quality factor $Q\approx$ 100. The qubit state is controlled by applying a magnetic flux in its ring $\Phi_{qb}(t)=\Phi_{qb,bias}+M_{control}I_{control}(t)$. $\Phi_{qb,bias}$ is generated using an external coil and set to a value close to one of the operation points $(n+1/2)\Phi_{0}$. $M_{control}I_{control}(t)$ is generated by an on-chip line and used to change the qubit state either adiabatically or by resonant pulses.  The qubit, resonant circuit, and control line are fabricated on an oxidized silicon substrate by using standard electron beam lithography. Measurements are performed in a dilution refrigerator, at temperatures of $\approx$~30~mK. 

The Josephson inductance of the SQUID depends on the current in a nonlinear way. Our resonant circuit is described, to a good level of approximation, by the Duffing oscillator model~\cite{dykman_1980_1}. This model predicts that a certain driving condition of the resonator (given by the driving frequency $F$ and amplitude $I_{r0}$) results in one unique forced oscillation state when $F>F_{res}(1-\sqrt{3}/(2Q))$. When $F<F_{res}(1-\sqrt{3}/(2Q))$, two oscillations states are possible for $I_{Bl}<I_{r0}<I_{Bh}$, and only one state otherwise. $I_{Bl}$ and $I_{Bh}$ are the values of the driving current where bifurcation occurs, dependent on $F$ and the parameters of the resonant circuit. The oscillation state denoted by 0 exists when $I_{r0}<I_{Bh}$, and the state 1 when $I_{r0}>I_{Bl}$. Without fluctuations the resonator, driven at frequency F and amplitude $I_{r0}$ slowly increasing from $I_{r0}=$0, resides in state 0 until $I_{r0}$ reaches $I_{Bh}$, when it switches to state 1. Thermal fluctuations will cause the transition $0\rightarrow 1$ to occur randomly at $I_{r0}<I_{Bh}$, with a rate that increases exponentially as $I_{r0}$ approaches $I_{Bh}$. The \emph{bistable} measurement regime relies on the fact that the bifurcation current $I_{Bh}$ depends on the value of the magnetic flux in the SQUID loop $\Phi_{sq}$. For a fixed driving condition of the oscillator, the probability for the transition $0\rightarrow 1$ to take place is very sensitive to variations in $\Phi_{sq}$, and thus to changes in the qubit state. The \emph{monostable} measurement regime does not exploit the $0\rightarrow 1$ transition, but relies instead on the measurement of the voltage and phase of the driven resonator, which also depend on $\Phi_{sq}$.

\begin{figure}[!]
\includegraphics[width=3.4 in]{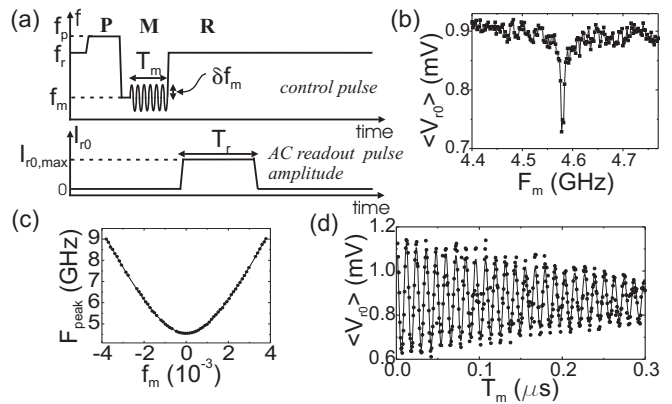}
\caption{\label{fig2} Measurements in the monostable regime. (a) Schematic representation of the qubit control (top) and readout (bottom) sequence; explanations are given in the text. (b) Spectroscopy measurement for $f_{m}=$ 0, $f_{p}=f_{r}=$ 0.0021, $T_{m}=$ 2.4 $\mu$s, and $T_{r}=$ 6 $\mu$s. (c) Plot of the qubit transition frequency, measured spectroscopically, versus the frustration $f_{m}$. The solid line is a fit that yields the qubit parameters, $\Delta$ and $I_{p}$, given in the text. (d) Measurement of Rabi oscillations with the same parameters as in (c) and $F_{m}=$ 4.602 GHz, resonant to the qubit energy level splitting. The solid line is a fit with an exponentially damped sinusoidal.}
\end{figure}

We present our experimental results starting with measurements in the \emph{monostable} regime. We control the qubit state by using a pulse as shown in Fig.~\ref{fig2}a (top). We start with a \emph{preparation} part (P), by fixing the frustration $f$ to a value $f_{p}$, for a time much longer than the qubit energy relaxation time. This results in a Boltzmann distribution for the qubit initial state, set by the effective temperature $T_{qb, eff}$. The qubit is prepared in the ground state if $T_{qb, eff} \ll E_{e}-E_{g}$. After qubit state preparation, the frustration $f$ is changed adiabatically to the value $f_{m}$, where qubit \emph{manipulation} (M) is done by using a microwave burst of frequency $F_{m}$ and duration $T_{m}$. Finally, the frustration $f$ is changed adiabatically to the value $f_{r}$ (for \emph{readout} - R), preserving the weight of the energy eigenstates created at M. Readout is possible, provided $\epsilon (f_{r}) \gtrsim \Delta$, because the two qubit states have significantly different currents (see Fig.~\ref{fig1}b).  To measure the state of the qubit, a microwave burst is sent to the resonator with frequency $F$, amplitude $I_{r0, max}$, and duration $T_{r}$ (see Fig.~\ref{fig2}a-bottom). The sequence shown in Fig.~\ref{fig2}a is repeated typically 10$^{5}$ times, and the average of the voltage of the transmitted wave $\langle V_{r0}\rangle$, which is a measure of qubit population, is calculated. Figure~\ref{fig2}b shows the result of a typical spectroscopy measurement. The position of the qubit spectroscopy peaks, $F_{peak}$, is plotted as a function of the frustration $f_{m}$ in Fig.~\ref{fig2}c. A fit of the data with the expectation for the qubit energy level splitting following from Eq.~\ref{eq_Ham}, yields the qubit parameters $\Delta=4.58$~GHz and $I_{p}=360$~nA. Using strong microwave pulses, with a frequency $F_{m}$ equal to the energy levels splitting of the qubit, we induce Rabi oscillations between the qubit energy eigenstates. A typical result is shown in Fig.~\ref{fig2}d, together with a fit with a exponentially damped sinusoidal signal. We observe Rabi oscillations with frequencies as high as 100 MHz, varying linearly with the amplitude $\delta f_{m}$ of the microwave burst.

\begin{figure}[!]
\includegraphics[width=3.4 in]{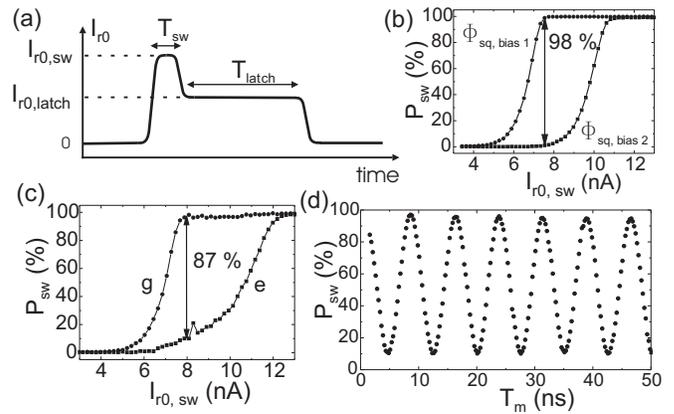}
\caption{\label{fig3} Measurements in the bistable regime. (a) Envelope of the AC readout pulse. (b) Switching probability curves for two values of $\Phi_{sq, bias}$: $\Phi_{sq, bias 1}=$ 2.335 $\Phi_{0}$ and $\Phi_{sq, bias 2}=$ 2.340 $\Phi_{0}$. (c) Switching probability curves, for qubit in the g and e states, for $f_{p}=$ 0.0161, $f_{m}=$ 0.00, $f_{r}=$ 0.0126, $T_{sw}\approx$ 100 ns, and $T_{latch}\sim$ 10 $\mu$s. (d) Rabi oscillation with the same setting as for (c), and $I_{sw}$ optimized for maximum contrast.}
\end{figure}

We now turn to a presentation of the experiments using the \emph{bistable} operation mode. To characterize experimentally the transition between the states 0 and 1 of the driven oscillator, we use the methods developed in~\cite{siddiqi_2004_1}. We drive the resonator with an AC pulse of fixed frequency $F<F_{res}(1-\sqrt{3}/(2Q))$ and modulation of the amplitude as shown in Fig.~\ref{fig3}a. During the first part of the pulse, of duration $T_{sw}$, the amplitude is set to a value $I_{r0, sw}$ close to $I_{Bh}$, such that there is a significant probability for the $0\rightarrow 1$ transition of the oscillator. The second part of the pulse is used to discriminate the states 0 and 1 by measuring the amplitude and phase of the voltage of the resonator. In this part $I_{r0}$ is set such that both transitions $0\rightarrow 1$ and $1\rightarrow 0$ have negligible rates and a long time $T_{latch}$ allows for amplifier noise suppression. We first measure the probability $P_{sw}$ of switching between the states \emph{0} and \emph{1}, with the qubit in its ground state, for two values of the magnetic flux in the SQUID loop $\Phi_{sq, bias 1}$ and $\Phi_{sq, bias 2}$. Their difference equals the change in magnetic flux when the qubit state changes from g to e. The switching probability $P_{sw}$ is shown in Fig.~\ref{fig3}b as a function of the amplitude $I_{r0,sw}$. The figure shows that it is possible, in principle, to detect the qubit state with an efficiency as high as 98~$\%$.

Using the bistable operation mode, we next measure changes in the state of the qubit. We use a control pulse as shown in Fig.~\ref{fig2}a (top) and a readout pulse as shown in Fig.~\ref{fig3}a. The shape of the control pulse is optimized to achieve high fidelity for qubit state preparation. Figure~\ref{fig3}c shows the switching probability curves when the qubit is prepared in the g or e state by using a resonant microwave pulse at M (see Fig.~\ref{fig2}a). The maximum separation between the two curves is $87\:\%$. This is smaller than the value of $98\:\%$ corresponding to a measurement of a magnetic flux difference equal to the qubit signal (see Fig.~\ref{fig3}b). Detailed experimental studies indicate that the most important contribution to the observed loss of contrast is the adiabatic shift from M to R, accounting for $\approx 7\:\%$. The qubit relaxation during the measurement (but \emph{not} due to the measurement process) accounts for $\approx 2\:\%$. The remainder of $\approx 3\:\%$ occurs during the switching on of the detector. In Fig.~\ref{fig3}c we show the measurement of Rabi oscillations with measurement settings optimized for maximum contrast. Each measurement is repeated $\sim 10^{4}$ times in order to reduce the spread of the average when measuring superpositions.

\begin{figure}[!]
\includegraphics[width=3.4 in]{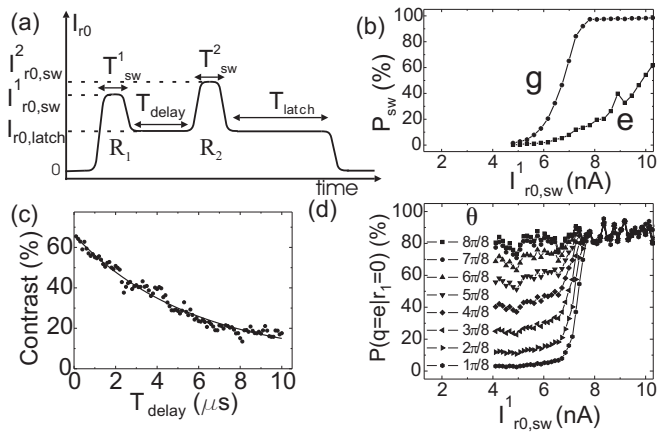}
\caption{\label{fig4}Results of the correlation measurements. (a) Envelope of the double-measurement pulse. (b) The switching probability curves for the qubit in the ground or excites states, with a single-measurement pulse with $T_{sw}^{1}\approx$ 125 ns. (c) The contrast of Rabi oscillations measured with a pulse as shown in (a), with $I_{r, sw}^{1}=I_{r, sw}^{2}=$7.7 nA and $T_{sw}^{1}=T_{sw}^{2}\approx$ 125 ns, as a function of $T_{delay}$. The solid line is a fit with an exponential decay. (d) Conditional probability for the qubit to be in the excited state, when the first measurement did not lead to switching, for different qubit initial states; $T_{delay}=800$ ns.}
\end{figure}

We finally address the question of how the qubit state is changed by the measurement~\cite{braginsky_1992_1}. This requires the use of two consecutive measurements; the second measurement is used to characterize the effect that the first measurement had on the qubit. Using two consecutive readout pulses, of the type shown in Fig.~\ref{fig3}a, has the disadvantage that the minimum duration of the latching plateau, due to the noise of our amplifiers, is comparable to the qubit relaxation time. The qubit would undergo significant energy relaxation before the second measurement takes place. We choose instead a readout pulse as shown in Fig.~\ref{fig4}a, consisting of two intervals of large amplitude, realizing two consecutive measurements R$_{1}$ and R$_{2}$. We use the labels $r_{1}$ and $r_{2}$ to denote the outcome of the measurements R$_{1}$ and R$_{2}$; they take the values $0$ ($1$) if switching did not (did) occur. For this double-measurement pulse the probability that switching occurs is $P_{sw}^{2}(|\psi\rangle_{init}, I_{r0,sw}^{1}, I_{r0,sw}^{2}, T_{delay})=1-P(r_{1}=0 \wedge r_{2}=0)$; here $|\psi\rangle_{init}$ is the initial qubit state, $I_{r0,sw}^{1}$ and $I_{r0,sw}^{2}$ are the amplitudes for R$_{1}$ and R$_{2}$, $T_{delay}$ is the delay between the two measurements, and $P(r_{1}=0\wedge r_{2}=0)$ is the probability that switching does not occur during either R$_{1}$ or R$_{2}$. We also read out the qubit state with a single measurement pulse by setting the amplitude $I_{r0,sw}^{2}=I_{r0,latch}$. The switching probability for this single-measurement pulse is $P_{sw}^{1}(|\psi\rangle_{init}, I_{r0,sw}^{1})=1-P(r_{1}=0)$, where $P(r_{1}=0)$ is the probability that switching does not occur during the measurement R$_{1}$. Figure~\ref{fig4}b shows the switching probability curves, for a single-measurement pulse, as a function of $I_{r0,sw}^{1}$. (The timing for this single-measurement pulse is as used in the experiments described below, but different from those that led to the data shown in Fig.~\ref{fig3}c).

In a first experiment we measure the contrast of Rabi oscillations with a double-measurement pulse as a function of the delay time $T_{delay}$ between R$_{1}$ and R$_{2}$, as shown in Fig.~\ref{fig4}c. The observed decrease is due to the relaxation of the qubit between R$_{1}$ and R$_{2}$. A fit with an exponential decay, shown by the solid line in Fig.~\ref{fig4}c, yields a decay time which is the same as the qubit relaxation time. This shows that the correlations between the results of measurements R$_{1}$ and R$_{2}$ are due to the change in the qubit state, and not to the detector dynamics.

In a second experiment we set the initial state of the qubit to a defined value and we measure the joint probability $P(r_{1}=0\wedge r_{2}=0)$ with a double-measurement pulse and the probability $P(r_{1}=0)$ with a single-measurement pulse. This allows calculating the probability for $r_{2}=0$ when $r_{1}=0$. Therefore the conditional probability $P(q=e|r_{1}=0)$, for the qubit to be in state $e$ when the first measurement gave the result $r_{1}=0$, can be calculated. This is plotted in Fig.~\ref{fig4}d versus $I_{r0,sw}^{1}$, for initial qubit states given by $|\psi\rangle_{init}=\cos(\theta/2)|g\rangle+\exp(i\phi)\sin(\theta/2)|e\rangle$ obtained by Rabi rotation; $\theta$ is indicated in Fig.~\ref{fig4}d for each curve, and $\phi$ is an unimportant phase angle. For large $I_{r0,sw}^{1}$ all the curves collapse to a single value of about $\approx 90\:\%$. A no-switch event results in a post-measurement state nearly equal to the excited state, \emph{irrespective of the pre-measurement state}. The region over which the curves in Fig.~\ref{fig4}d show a fast increase corresponds to the region where the switching probability for the ground state, shown in Fig.~\ref{fig4}b, increases; this can be interpreted as a transition from weak to strong measurement. Note that using the double-measurement pulse shown in Fig.~\ref{fig4}a instead of two successive pulses as shown in Fig.~\ref{fig3}a has the limitation of only giving information on the qubit state when $r_{1}=0$.

In this paper we presented measurements of a superconducting qubit using a dispersive readout scheme. Operation in the bistable mode results in the observation of Rabi oscillations with very high contrast, significantly improved over other measurements of this type. We also present measurements which indicate that a single measurement prepares the qubit state with a fidelity of about 90~$\%$. These results establish dispersive readout as very suited for flux qubits and make it promising for readout of entangled two-qubit states.

We acknowledge the help of P. Bertet and R.N. Schouten and useful discussions with I. Siddiqi, R. Vijay, and L. Vandersypen. This work was supported by the SQUBIT project and the Large Scale Installation Program ULTI-3 of the European Union, the Dutch Organization for Fundamental Research on Matter (FOM), the Academy of Finland, and the NanoNed program.


\begin{thebibliography}{20}
\expandafter\ifx\csname natexlab\endcsname\relax\def\natexlab#1{#1}\fi
\expandafter\ifx\csname bibnamefont\endcsname\relax
  \def\bibnamefont#1{#1}\fi
\expandafter\ifx\csname bibfnamefont\endcsname\relax
  \def\bibfnamefont#1{#1}\fi
\expandafter\ifx\csname citenamefont\endcsname\relax
  \def\citenamefont#1{#1}\fi
\expandafter\ifx\csname url\endcsname\relax
  \def\url#1{\texttt{#1}}\fi
\expandafter\ifx\csname urlprefix\endcsname\relax\def\urlprefix{URL }\fi
\providecommand{\bibinfo}[2]{#2}
\providecommand{\eprint}[2][]{\url{#2}}

\bibitem[{\citenamefont{Devoret et~al.}()\citenamefont{Devoret, Wallraff, and
  Martinis}}]{_devoret_2004_1}
\bibinfo{author}{\bibfnamefont{M.}~\bibnamefont{Devoret}},
  \bibinfo{author}{\bibfnamefont{A.}~\bibnamefont{Wallraff}}, \bibnamefont{and}
  \bibinfo{author}{\bibfnamefont{J.}~\bibnamefont{Martinis}},
  \bibinfo{journal}{cond-mat/0411174}.

\bibitem[{\citenamefont{Nielsen and Chuang}(2000)}]{nielsen_2000_1}
\bibinfo{author}{\bibfnamefont{M.~A.} \bibnamefont{Nielsen}} \bibnamefont{and}
  \bibinfo{author}{\bibfnamefont{I.~L.} \bibnamefont{Chuang}},
  \emph{\bibinfo{title}{Quantum Computation and Quantum Information}}
  (\bibinfo{publisher}{Cambridge University Press}, \bibinfo{year}{2000}).

\bibitem[{\citenamefont{Collin et~al.}(2004)\citenamefont{Collin, Ithier,
  Aassime, Joyez, Vion, and Esteve}}]{collin_2004_1}  
\bibinfo{author}{\bibfnamefont{E.}~\bibnamefont{Collin}} \textit{et al.},
  \bibinfo{journal}{Phys. Rev. Lett.} \textbf{\bibinfo{volume}{93}},
  \bibinfo{pages}{157005} (\bibinfo{year}{2004}).

\bibitem[{\citenamefont{Yamamoto et~al.}(2003)\citenamefont{Yamamoto, Pashkin,
  Astafiev, Nakamura, and Tsai}}]{yamamoto_2003_1}
\bibinfo{author}{\bibfnamefont{T.}~\bibnamefont{Yamamoto}} \textit{et al.},
  \bibinfo{journal}{Nature} \textbf{\bibinfo{volume}{425}},
  \bibinfo{pages}{941} (\bibinfo{year}{2003}).

\bibitem[{\citenamefont{McDermott et~al.}(2003)\citenamefont{McDermott,
  Simmonds, Steffen, Cooper, Cicak, Osborn, Oh, Pappas, and
  Martinis}}]{mcdermott_2005_1}
\bibinfo{author}{\bibfnamefont{R.}~\bibnamefont{McDermott}} \textit{et al.},
  \bibinfo{journal}{Science} \textbf{\bibinfo{volume}{307}},
  \bibinfo{pages}{1299} (\bibinfo{year}{2003}).

\bibitem[{\citenamefont{Vion et~al.}(2002)\citenamefont{Vion, Aassime, Cottet,
  Joyez, Pothier, Urbina, Esteve, and Devoret}}]{vion_2002_1}
\bibinfo{author}{\bibfnamefont{D.}~\bibnamefont{Vion}} \textit{et al.},
  \bibinfo{journal}{Science} \textbf{\bibinfo{volume}{296}},
  \bibinfo{pages}{886} (\bibinfo{year}{2002}).

\bibitem[{\citenamefont{Chiorescu et~al.}(2003)\citenamefont{Chiorescu,
  Nakamura, Harmans, and Mooij}}]{chiorescu_2003_1}
\bibinfo{author}{\bibfnamefont{I.}~\bibnamefont{Chiorescu}} \textit{et al.},
  \bibinfo{journal}{Science} \textbf{\bibinfo{volume}{299}},
  \bibinfo{pages}{1869} (\bibinfo{year}{2003}).

\bibitem[{\citenamefont{Duty et~al.}(2004)\citenamefont{Duty, Gunnarsson,
  Bladh, and Delsing}}]{duty_2004_1}
\bibinfo{author}{\bibfnamefont{T.}~\bibnamefont{Duty}} \textit{et al.},
  \bibinfo{journal}{Phys. Rev. B} \textbf{\bibinfo{volume}{69}},
  \bibinfo{pages}{140503} (\bibinfo{year}{2004}).

\bibitem[{\citenamefont{Astafiev et~al.}(2004)\citenamefont{Astafiev, Pashkin,
  Yamamoto, Nakamura, and Tsai}}]{astafiev_2004_1}
\bibinfo{author}{\bibfnamefont{O.}~\bibnamefont{Astafiev}} \textit{et al.},
  \bibinfo{journal}{Phys. Rev. B} \textbf{\bibinfo{volume}{69}},
  \bibinfo{pages}{180507} (\bibinfo{year}{2004}).

\bibitem[{\citenamefont{Cooper et~al.}(2004)\citenamefont{Cooper, Steffen,
  McDermott, Simmonds, Oh, Hite, Pappas, and Martinis1}}]{cooper_2004_1}
\bibinfo{author}{\bibfnamefont{K.~B.} \bibnamefont{Cooper}} \textit{et al.},
  \bibinfo{journal}{Phys. Rev. Lett.}
  \textbf{\bibinfo{volume}{93}}, \bibinfo{pages}{180401}
  (\bibinfo{year}{2004}).

\bibitem[{\citenamefont{Grajcar et~al.}(2004)\citenamefont{Grajcar, Izmalkov,
  Il'ichev, Wagner, Oukhanski, Hubner, May, Zhilyaev, Hoenig, Greenberg
  et~al.}}]{grajcar_2004_1}
\bibinfo{author}{\bibfnamefont{M.}~\bibnamefont{Grajcar}} \textit{et al.},
  \bibinfo{journal}{Phys. Rev. B}
  \textbf{\bibinfo{volume}{69}}, \bibinfo{pages}{60501} (\bibinfo{year}{2004}).

\bibitem[{\citenamefont{Lupa\c{s}cu et~al.}(2004)\citenamefont{Lupa\c{s}cu,
  Verwijs, Schouten, Harmans, and Mooij}}]{lupascu_2004_1}
\bibinfo{author}{\bibfnamefont{A.}~\bibnamefont{Lupa\c{s}cu}} \textit{et al.},
  \bibinfo{journal}{Phys. Rev. Lett.} \textbf{\bibinfo{volume}{93}},
  \bibinfo{pages}{177006} (\bibinfo{year}{2004}).

\bibitem[{\citenamefont{Wallraff et~al.}(2005)\citenamefont{Wallraff, Schuster,
  Blais, Frunzio, Majer, Devoret, Girvin, and Schoelkopf}}]{wallraff_2005_1}
\bibinfo{author}{\bibfnamefont{A.}~\bibnamefont{Wallraff}} \textit{et al.},
  \bibinfo{journal}{Phys. Rev. Lett.} \textbf{\bibinfo{volume}{95}},
  \bibinfo{pages}{060501} (\bibinfo{year}{2005}).

\bibitem[{\citenamefont{Siddiqi et~al.}()\citenamefont{Siddiqi, Vijay,
  Metcalfe, Boaknin, Frunzio, Schoelkopf, and Devoret}}]{_siddiqi_2005_1}
\bibinfo{author}{\bibfnamefont{I.}~\bibnamefont{Siddiqi}} \textit{et al.},
  \bibinfo{journal}{cond-mat/0507548}.

\bibitem[{\citenamefont{Sillanpaa et~al.}(2004)\citenamefont{Sillanpaa, Lehtinen,
  Paila, Makhlin, Roschier, and Hakonen}}]{sillanpaa_2005_1}
\bibinfo{author}{\bibfnamefont{M.A.}~\bibnamefont{Sillanp\"{a}\"{a}}} \textit{et al.},
  \bibinfo{journal}{Phys. Rev. Lett.} \textbf{\bibinfo{volume}{95}},
  \bibinfo{pages}{206806} (\bibinfo{year}{2005}).

\bibitem[{\citenamefont{Siddiqi et~al.}(2004)\citenamefont{Siddiqi, Vijay,
  Pierre, Wilson, Metcalfe, Rigetti, Frunzio, and Devoret}}]{siddiqi_2004_1}
\bibinfo{author}{\bibfnamefont{I.}~\bibnamefont{Siddiqi}} \textit{et al.},
  \bibinfo{journal}{Phys. Rev. Lett.} \textbf{\bibinfo{volume}{93}},
  \bibinfo{pages}{207002} (\bibinfo{year}{2004}).

\bibitem[{\citenamefont{Mooij et~al.}(1999)\citenamefont{Mooij, Orlando,
  Levitov, Tian, van~der Wal, , and Lloyd}}]{mooij_1999_1}
\bibinfo{author}{\bibfnamefont{J.~E.} \bibnamefont{Mooij}} \textit{et al.},
  \bibinfo{journal}{Science} \textbf{\bibinfo{volume}{285}},
  \bibinfo{pages}{1036} (\bibinfo{year}{1999}).

\bibitem[{\citenamefont{Barone and Paterno}(1982)}]{barone_1982_1}
\bibinfo{author}{\bibfnamefont{A.}~\bibnamefont{Barone}} \bibnamefont{and}
  \bibinfo{author}{\bibfnamefont{G.}~\bibnamefont{Paterno}},
  \emph{\bibinfo{title}{Physics and Applications of the Josephson effect}}
  (\bibinfo{publisher}{John Wiley and Sons}, \bibinfo{year}{1982}).

\bibitem[{\citenamefont{Roschier and Hakonen}(2004)}]{roschier_2004_1}
\bibinfo{author}{\bibfnamefont{L.}~\bibnamefont{Roschier}} \bibnamefont{and}
  \bibinfo{author}{\bibfnamefont{P.}~\bibnamefont{Hakonen}},
  \bibinfo{journal}{Cryogenics} \textbf{\bibinfo{volume}{44}},
  \bibinfo{pages}{783} (\bibinfo{year}{2004}).

\bibitem[{\citenamefont{Dykman and Krivoglaz}(1980)}]{dykman_1980_1}
\bibinfo{author}{\bibfnamefont{M.}~\bibnamefont{Dykman}} \bibnamefont{and}
  \bibinfo{author}{\bibfnamefont{M.}~\bibnamefont{Krivoglaz}},
  \bibinfo{journal}{Physica A} \textbf{\bibinfo{volume}{104}},
  \bibinfo{pages}{480} (\bibinfo{year}{1980}).

\bibitem[{\citenamefont{Braginsky and Khalili}(1992)}]{braginsky_1992_1}
\bibinfo{author}{\bibfnamefont{V.}~\bibnamefont{Braginsky}} \bibnamefont{and}
  \bibinfo{author}{\bibfnamefont{F.}~\bibnamefont{Khalili}},
  \emph{\bibinfo{title}{Quantum Measurement}} (\bibinfo{publisher}{Cambridge
  University Press}, \bibinfo{address}{Cambridge}, \bibinfo{year}{1992}).

\end{thebibliography}

\end{document}